\documentclass[conference]{IEEEtran}

\usepackage{alltt                                    
          , multirow
          , booktabs
          , listings
          , graphicx
          ,float
	,cite
          ,verbatim
         ,mathtools
	,url
	,amsmath
}
\usepackage[table]{xcolor}
\usepackage[numbers]{natbib}     
\usepackage{syntax}
\usepackage{algorithmic, algorithm}
\usepackage{enumitem}
\usepackage{threeparttable}
\usepackage{framed}
\usepackage{hhline}

\usepackage{expl3}
\ExplSyntaxOn
\newcommand\latinabbrev[1]{
  \peek_meaning:NTF . {
    #1\@}%
  { \peek_catcode:NTF a {
      #1., \@ }%
    {#1., \@}}}
\ExplSyntaxOff

\newcommand{\CASE}[1]{\STATE \textbf{case} #1\textbf{:} \begin{ALC@g}}
\newcommand{\ENDCASE}{\end{ALC@g}}

\newcommand{\DEFAULT}{\STATE \textbf{default:} \begin{ALC@g}}
\newcommand{\ENDDEFAULT}{\end{ALC@g}}
\newcommand{\DEFAULTLINE}[1]{\STATE \textbf{default:} }
\let\footnotesize\scriptsize

\newsavebox{\supbox}
\newcommand{\bsup}{\begin{lrbox}{\supbox}$\tt\scriptstyle}
\newcommand{\esup}{$\end{lrbox}{}^{\usebox{\supbox}}}
\def\eg{\latinabbrev{e.g}}
\def\ie{\latinabbrev{i.e}}

\definecolor{lightpurple}{rgb}{0.8,0.8,1}
\definecolor{codebg}{RGB}{255,255,255}
\definecolor{commentcolor}{RGB}{11,140,11}
\begin{document}
%

\title{TextRank Based Search Term Identification for Software Change Tasks\vspace{-.4cm}}
%
%
%
%
%

\author{\IEEEauthorblockN{Mohammad Masudur Rahman  ~~~ Chanchal K. Roy}
\IEEEauthorblockA{University of Saskatchewan, Canada\\
\{masud.rahman, chanchal.roy\}@usask.ca}
}

\maketitle

\begin{abstract}
During maintenance, software developers deal with a number of software change requests. 
Each of those requests is generally written using natural language texts, and it involves one or more domain related concepts. A developer needs to map those concepts to exact source code locations within the project in order to implement the requested change.
This mapping generally starts with a search within the project that requires one or more suitable search terms.
Studies suggest that the developers often perform poorly in coming up with good search terms for a change task. 
In this paper, we propose and evaluate a novel \emph{TextRank}-based technique that automatically identifies and suggests search terms for a software change task by analyzing its task description.
Experiments with 349 change tasks from two subject systems and comparison with one of the latest and closely related state-of-the-art approaches show that our technique is highly promising in terms of \emph{suggestion accuracy}, \emph{mean average precision} and \emph{recall}.

\end{abstract}


\IEEEpeerreviewmaketitle

\section{Introduction}
Studies show that about 80\% of the total efforts is spent in software maintenance and evolution \cite{seahawk}. During maintenance, software developers deal with a number of software change requests, and identifying the exact location (\ie\ class, method) within a project for a given change task is a major challenge. Change requests are often made by software users, and are generally written using natural language texts. The software users might be familiar with the application domain of a software product; however, they generally lack the idea of how a particular \emph{software feature} is implemented in the source code. Thus a \emph{software change request} by them mainly involves one or more domain related concepts, and a developer needs to map those concepts to the source code locations within the project in order to implement the change \cite{kevicdict, textret}. Such mapping is possibly trivial for a developer who has substantial knowledge about the target project. However, the developers involved in maintenance might be unaware of the low-level architecture of the project, and thus they often experience difficulties in identifying the source locations (\ie\ classes, methods) to be changed. The mapping task generally starts with a search within the project which requires one or more suitable search terms. 
Previous studies \cite{kevic} suggest that on average, developers with varying experience perform poorly in coming up with good search terms for a change task. 
For example, according to \citet{kevic}, developers can come up with relevant search terms for only 12.2\% of the cases.
One way to help them in this regard is to automatically suggest useful and relevant search terms for the change task in the first place.

Existing studies that attempt to support developers in \emph{feature location} tasks with search queries, adopt different lightweight heuristics \cite{kevic} and query reformulation or expansion strategies \cite{gayg, refoqus, shepherd}, and perform different query quality analyses \cite{qquality,qeffect,specificity} and mining activities \cite{kevicdict,ccmapping}. However, most of these approaches expect a developer to provide the initial search query which they can improve upon, and it is often a non-trivial task for the developers as noted by other studies too \cite{kevic}. \citet{kevic} propose a heuristic model for automatically identifying initial search terms for a change task where they consider different heuristics related to \emph{frequency, location, parts of speech} and \emph{notation} of the terms from the task description. Although the model is found promising in their preliminary evaluation, it suffers from two major limitations. First, the model is trained using a limited number of change tasks, and is not cross-validated using the change tasks from another project. Thus it is still not quite mature or reliable. Second, \emph{tf--idf} is one of the dominating metrics in their model, and it is subject to the size of test dataset for \emph{inverse document frequency (idf)} calculation. Thus the same model is likely to perform differently with different sizes of test dataset, and the model might require frequent re-training to keep itself useful. 

In this paper, we propose a novel \emph{TextRank}-based technique that automatically identifies and suggests search terms for software change tasks. \emph{TextRank} is an adaptation of \emph{PageRank algorithm} \cite{pagerank} for natural language texts where a text document is considered as a lexical or semantic network of words \cite{rada, blanco}. \emph{TextRank} has found its applications in different information retrieval tasks such as keyword and key phase extraction, extractive summarization, word sense disambiguation and other tasks involving \emph{graph-based} term weighting \cite{rada}.
Given the successful adoption of that algorithm in information retrieval domain, it can also be exploited for search term extraction in the context of \emph{feature location} tasks. Actually, the algorithm is highly suited for our purpose from several perspectives.
First, software change requests are generally made by the people outside the development team, and they communicate their requirements through domain level concepts and using natural language texts. A \emph{graph-based} representation (\ie\ also called \emph{text graph}) of the task description can reveal  important semantic (\ie\ co-occurrence) relationships among different terms. Second, \emph{TextRank} algorithm exploits connectivity of a term in the graph, and considers not only the terms to which the term is connected but also their weights (\ie\ importance) for determining the weight of that term. The algorithm continues this process recursively, and thus has a
great potential to extract the most important terms from the 
graph which can be suggested as search terms.

\FrameSep5pt
\begin{framed}\label{frame:ctask}
\noindent
\textbf{Issue ID}: 401358\\
\textbf{Product:} JDT, \textbf{Component:} Debug\\
\textbf{Summary}: Name \emph{selection} for \emph{Mac} VM \emph{installs} needs \emph{improvement} \\
\textbf{Description}: When you search for a JDK/JRE on \emph{Mac}, we use information from the plist file to compute a name. This works fine most of the time, but if you happen to have more than one of the same version of VM installed they are added with the same name. To make matters a bit worse, if you edit one of the \emph{JREs} the wizard starts out with an error complaining that the name is already in use. The attached screen shot shows the duplicated names for the Java 7 \emph{JREs}.
\end{framed}
\begin{center}
\selectfont\scriptsize
Listing 1: An Example change request from \texttt{eclipse.jdt.debug}\\
\end{center}

\begin{figure}[!t]
\centering
\includegraphics[width=2.5in]{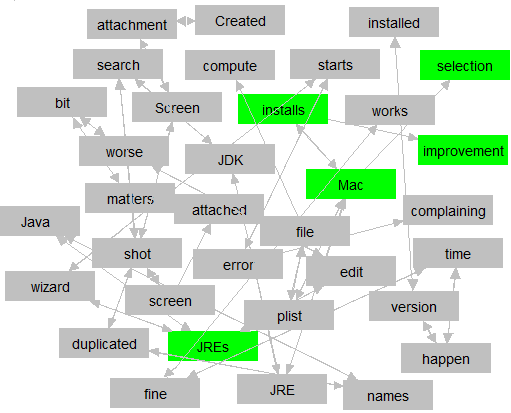}
\vspace{-.2cm}
\caption{Text Graph of change request in Listing 1}
\vspace{-.4cm}
\label{fig:tgraph}
\end{figure}

\noindent
For example, 
the change task in Listing 1 can be graphically represented  as the \emph{text graph} in Fig. \ref{fig:tgraph} by analyzing the \emph{co-occurrence relationships} among the words with a \emph{window size} (\ie\ number of words considered in a semantic text unit) of two. 
Our proposed technique analyzes the term connectivity, calculates weight (\ie\ importance) for each of the terms in the graph, and then extracts the top-scored and the most suitable five terms-- \emph{Mac, selection, installs, improvement} and \emph{JREs} as the initial search query for the task.

Experiments with 349 change tasks from two  subject systems-- \emph{Apache Log4j} and \emph{Eclipse JDT Debug} show that our approach can return relevant results (\ie\ Java classes) for 49.34\% of the change tasks with 57.16\% \emph{mean average precision} and 63.33\% \emph{recall}.
We also compare with one of the latest and closely related state-of-the-art approaches-- \citet{kevic}, and report that our approach performs relatively better in terms of all performance metrics.
While our preliminary results are found promising, they must be further validated with more change tasks from more subject systems. Thus the paper makes the following technical contributions:
\begin{itemize}
\item It demonstrates a novel use of \emph{TextRank} algorithm in the identification and suggestion of relevant search terms for software change tasks. 
\item It reports a case study involving 349 software change tasks from two subject systems and comparison with one existing approach \cite{kevic}, and shows the effectiveness of the proposed technique.
\end{itemize}

The rest of the paper is organized as follows -- Section \ref{sec:proposed} explains our adopted methodology and algorithms, Section \ref{sec:experiment} discusses the conducted experiments and findings, Section \ref{sec:related} describes the related work,  and finally Section \ref{sec:conclusion} concludes the paper with future work.


\begin{figure*}[!t]
\centering
\includegraphics[width=4in]{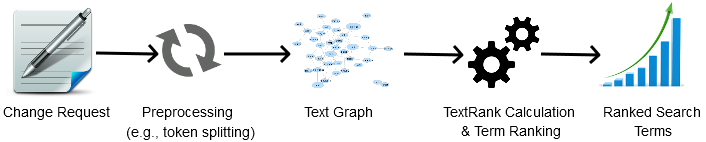}
\vspace{-.2cm}
\caption{Schematic diagram of the proposed technique}
\vspace{-.4cm}
\label{fig:sysdiag}
\end{figure*}

\vspace{-.2cm}

\section{Proposed Methodology}
\label{sec:proposed}
Fig. \ref{fig:sysdiag} shows the schematic diagram of our proposed technique for search term identification and suggestion for a change task. In this section, we discuss 
the different steps involved with the technique as follows:

\subsection{Data Collection}\label{sec:datacoll}
In order to suggest search terms for a change task, we make use of natural language description (\eg\ Listing 1) for the task provided by the user. We collect actual task descriptions of 349 tasks from BugZilla official repositories.
Each of those tasks is submitted using a semi-structured way, and they contain several fields such as \emph{Issue ID} (\eg\ 401358), \emph{Product} (\eg\ JDT), \emph{Component} (\eg\ Debug), \emph{Summary} and \emph{Description}. We use the last two fields for the analysis by our technique.
While \emph{Summary} shows the title of a requested change task, \emph{Description} contains the user's  explanation of the task in natural language texts.
In order to keep the algorithm simple and lightweight, we do not consider content from the files attached to the change request.
Adding more available information about the task is likely to improve the performance of our technique; however, we consider that part as a scope of future study. 

\subsection{Text Preprocessing}\label{sec:preprocess}
We analyze \emph{Summary} and \emph{Description} of a software change request, and perform several preprocessing before transforming the texts into a \emph{text graph}. We consider each sentence as a \emph{logical text unit} that contributes to the overall task description, and collect each of them from those fields. 
We attempt to extract such terms from a sentence that either convey useful semantics or represent the domain level concepts of the software to be changed. 
We thus remove stop words (\ie\ frequently used non-important words) from each of those sentences, and split the dotted structured words (\ie\ containing dots) into simpler words from them. 
A dotted word often involves multiple technical concepts, and splitting helps one to analyze each of them in isolation. For example the word-- \texttt{org.eclipse.ui.part.PageBookView.create\\PartControl} contains a \emph{package name} (\ie\ \emph{org.eclipse.ui.part}), a \emph{class name} (\ie\ \emph{PageBookView}) and a \emph{method name} (\ie\ \emph{createPartControl}), and only splitting helps to analyze those concepts effectively.
It should be noted that we do not split the words based on camel-case notation given these two observations. First, change requests often contain different technical artifacts such as \emph{library name}, \emph{class name} and \emph{method name} using camel-case notations in the description, and they are of great interest to the developers implementing those changes. Splitting such words simply breaks those artifacts, and does not help in identifying suitable search terms. Second, experiments with such splitting do not result into performance improvement of our technique.
We also avoid stemming on the words of the task description as it does not demonstrate any notable improvement in the performance of our technique.

\subsection{Text Graph Development}\label{sec:tokengraph}
After preprocessing, we get a list of sentences each of which contains semantically important and domain related terms, and we use them to develop a \emph{term graph} for the task description.
We represent each of those terms as a distinct node in the graph, and consider the \emph{co-occurrence} of those terms in the sentences as an indication of semantic relationships (\ie\ dependencies) among them \cite{rada, blanco}.
For example, if we consider the sentence-- \emph{"This works fine most of the time, but if you happen to have more than one of the same version of VM installed they are added with the same name"} from the example description (Listing 1), the preprocessed version
forms an ordered list of terms-- \emph{"works fine time happen version installed"}. 
We note that the transformed sentence contains several phrases such as \emph{works fine} and \emph{version installed}, and the terms in those phrases are semantically dependent on each other for comprehensive meaning.
We thus consider a \emph{window size} of two (\ie\ best performing size according to \citet{rada}) as a semantic unit of words, and get the following relationships-- \emph{works}$\longleftrightarrow$\emph{fine}, \emph{fine}$\longleftrightarrow$\emph{time}, \emph{time}$\longleftrightarrow$\emph{happen}, \emph{happen}$\longleftrightarrow$\emph{version}, and \emph{version}$\longleftrightarrow$\emph{installed}. We then encode such relationships into the connecting edges among the corresponding nodes in the graph.

\subsection{TextRank Calculation}
\label{sec:textrank}
In order to estimate the weight (\ie\ importance) of each of the terms in \emph{text graph}, we use \emph{TextRank} algorithm by \citet{rada} which is adapted from the popular \emph{PageRank} algorithm \cite{pagerank} for web link analysis. \emph{TextRank} analyzes the connectivity details such as incoming links and outgoing links of each term in the graph recursively, and calculates its weight, $S(v_{i})$, as follows:
\begin{equation*}\label{eq:textrank}
\setlength\abovedisplayskip{1pt}
\setlength\belowdisplayskip{1pt}
S(v_{i})=(1-\phi)+\phi\sum_{j\epsilon In(v_{i})}\frac{S(v_{j})}{|Out(v_{j})|}~~ (0 \le \phi \le1)
\end{equation*}
\noindent
Here, $In(v_{i})$, $Out(v_{j})$, and $\phi$ denote the node list connected to $v_i$ through incoming links, the node list to which $v_j$ is connected through outgoing links, and the dumping factor respectively. In the \emph{text graph}, each of the edges is bi-directional (\ie\ terms depend on each other), and thus \emph{in-degree} is equal to the \emph{out-degree} for a node (\ie\ term).
In the context of web surfing, dumping factor, $\phi$, is considered as the probability of randomly choosing a web page by the surfer, and $1-\phi$ is the probability of jumping off that page. \citet{rada} use a heuristic value of 0.85 for $\phi$, and we also use the same value for \emph{TextRank} calculation. We initialize each of the terms in the graph with a default value of 0.25,
and run an iterative version of the algorithm \cite{pagerank}. It should be noted that the initial value of a term does not affect its final score \cite{rada}.
The computation iterates until the scores of the terms converge below a certain threshold or it reaches the maximum iteration limit (\ie\ 100 as suggested by \citet{blanco}). As \citet{rada} suggest, we use a heuristic threshold of 0.0001 for the convergence checking.
\emph{TextRank} applies the underlying mechanism of a recommendation (\ie\ voting) system, where a term recommends (\ie\ votes) another term if the second term complements the semantics of the first term in any way \cite{rada}. 
The algorithm captures recommendation for a term in terms of incoming links in the \emph{text graph} (\eg\ Fig. \ref{fig:tgraph}) from another terms both in local (\ie\ same sentence) and global (\ie\ entire document) context, and determines importance of the term.
Once computation is over, each of the terms in the graph is found with a score which is considered as the weight or importance of that term in the texts.

\subsection{Search Term Selection}\label{sec:searchterm}
Once \emph{TextRank} is calculated, we rank the terms based on their weights (\ie\ importance), and choose the search terms for a change task in a heuristic fashion. According to \citet{kevic}, the terms that exist both in \emph{Summary} and \emph{Description} of a task are the most suitable for search terms. 
We adapt that idea given the fact that the overlapping terms in those fields might not be sufficient enough to form a search query. We thus first look for top-scored five terms in the \emph{Summary} (\ie\ title) of a change task. If the \emph{Summary} is too small to provide all the terms, we collect the rest from the \emph{Description} of the task.
It should be noted that in both cases, terms are chosen based on their ranks or weights which are calculated by recursively analyzing the surrounding terms in the text graph.
For instance, in the case of our example change task (Listing 1), the first four search terms-- \emph{Mac (score: 0.64), selection (score: 0.27), installs (score: 0.27)} and \emph{improvement (score: 0.25)} come from the title. The remaining search term--\emph{JREs (score: 1.00)}, one of the most important terms, is not contained in the title, and thus it comes from the description.

\begin{table*}
\centering
\caption{Experimental Results}\label{table:result}
\resizebox{5.5in}{!}{%
\begin{threeparttable}
\begin{tabular}{l|c|c|c||c|c|c||c}
\hline
\multirow{2}{*}{\textbf{Metric}} & \multicolumn{3}{c||}{\texttt{Log4j}} & \multicolumn{3}{c||}{\texttt{eclipse.jdt.debug}} & \textbf{Average}\\
\hhline{~-------}
& $\mathbf{T_3}\tnote{1}$ & $\mathbf{T_4}\tnote{2}$ & $\mathbf{T_5}$\tnote{3} & $\mathbf{T_3}$ & $\mathbf{T_4}$ & $\mathbf{T_5}$ & $\mathbf{T_5}$  \\
\hline
No. of Tasks Solved (NTS) & 86(230) & 98(230) & \textbf{111}(230) & 48(119) & 54(119) & \textbf{60}(119) & -- \\
\hline
\% of Tasks Solved (PTS) & 37.39\% & 42.61\% & \textbf{48.26}\% & 40.34\% & 45.38\% & \textbf{50.42}\% & 49.34\%\\
\hline
Mean Average Precision (MAP) & \textbf{66.54}\% & 62.82\% & 61.16\% & \textbf{56.79}\% & 51.45\% & 53.16\% & 57.16\% \\
\hline
Mean Recall (MR) & 53.22\% & \textbf{55.54}\% & 54.66\% & \textbf{73.56}\% & 73.27\% & 72.00\% & 63.33\% \\
\hline
\end{tabular}
\centering
 $^1$Results for three search terms,  $^2$Results for four search terms,  $^3$Results for five search terms
\end{threeparttable}
}
\vspace{-.5cm}
\end{table*}

\section{Experiment}
\label{sec:experiment}
In order to evaluate our proposed technique, we conduct experiments using the change tasks from two subject systems. We also compare with one existing approach, and this section discusses our evaluation and validation details.

\subsection{Dataset}\label{sec:dataset}
In our experiments, we use 349 change tasks from two subject systems-- \texttt{Log4j} by \emph{Apache} and \texttt{eclipse.jdt.debug} by \emph{Eclipse}. Each of those tasks were marked as \emph{RESOLVED}, and we follow a careful approach for their selection. First we collect all the \emph{RESOLVED} change tasks from the BugZilla repositories \cite{log4j,debug}, and then attempt to map them against the commit history of the corresponding projects (\ie\ repositories) at GitHub \cite{log4jgh,debuggh}. We analyze the commit messages of each project, and look for specific \emph{Bug IDs} (\ie\ identifiers of change tasks) in those messages. In GitHub, we note that each commit operation that solves a software bug or addresses a change request, generally mentions the corresponding \emph{Bug ID} in the very first sentence of its commit message.
We found such 230 commits in \texttt{Log4j} and 119 commits in \texttt{eclipse.jdt.debug}. We then collect the \emph{changeset} (\ie\ list of changed files) for each of those commit operations, and develop a solution set for the corresponding change tasks. Thus, for our experiments, we collect not only the actual change tasks from the reputed subject systems but also their solutions that were applied in practice by the developers. We use different utility commands such as \emph{git, clone, rev-list} and \emph{log} on \emph{GitHub Bash} for collecting those information. 

\subsection{Search Engine}\label{sec:sengine}
We use a \emph{vector space model} based search engine-- \emph{Apache Lucene} \cite{seahawk} for searching the files that need to be changed for a change task. 
The search engine indexes the files in the corpus prior to search, and we note that the indexing by \emph{Lucene Indexer} is not efficient.
Especially the terms indexed from the source files are not meaningful and they often contain different special characters (\ie\ punctuations).
The indexer is basically targeted for simple text documents whereas the source files in the project contain items beyond regular texts (\eg\ code segments).
We thus apply limited preprocessing (\ie\ stemming was avoided) on each of those source files in each project, and remove all punctuation characters from them.
This transforms the source files into text like files, and the indexer becomes able to perform more effectively, especially in choosing meaningful index terms.  
Once a search is initiated using a query, the search engine filters irrelevant files in the corpus using a \emph{boolean search model}, and then applies a \emph{tf-idf} based scoring technique to return a ranked list of relevant documents.
As existing studies suggest \cite{antoniol, kevic, marcus}, we consider only the top ten results from the search engine for performance evaluation and validation of our technique.

\begin{table*}
\centering
\caption{Comparison with an Existing Approach}\label{table:comparison}
\resizebox{5in}{!}{%
\begin{threeparttable}
\begin{tabular}{l|l|c|c||c|c}
\hline
\multirow{2}{*}{\textbf{Technique}} & \multirow{2}{*}{\textbf{Metric}} & \multicolumn{2}{c||}{\texttt{Log4j}} & \multicolumn{2}{c}{\texttt{eclipse.jdt.debug}}\\
\hhline{~~----}
& & $\mathbf{T_3}\tnote{1}$ &  $\mathbf{T_5}$\tnote{2} & $\mathbf{T_3}$ & $\mathbf{T_5}$  \\
\hline
\multirow{4}{*}{\citet{kevic}} & No. of Tasks Solved (NTS) & 47(230)  & \textbf{65}(230) & 27(119) & \textbf{39}(119)  \\
\hhline{~-----}
& \% of Tasks Solved (PTS) & 20.43\% & 28.26\% & 22.69\% & 32.77\% \\
\hhline{~-----}
& Mean Average Precision (MAP) & 54.08\% & \textbf{56.90}\% & 50.61\% & \textbf{54.92}\%  \\
\hhline{~-----}
& Mean Recall (MR) & \textbf{50.39}\%  & 48.36\% & 66.70\%  & \textbf{78.53}\%  \\
\hline
\hline
\multirow{4}{*}{Proposed} & No. of Tasks Solved (NTS) & 86(230)  & \textbf{111}(230) & 48(119) & \textbf{60}(119)  \\
\hhline{~-----}
& \% of Tasks Solved (PTS) & 37.39\% & 48.26\% & 40.34\% & 50.42\% \\
\hhline{~-----}
& Mean Average Precision (MAP) & \textbf{66.54}\% & 61.16\% & \textbf{56.79}\% & 53.16\% \\
\hhline{~-----}
& Mean Recall (MR) & 53.22\%  & \textbf{54.66}\% & \textbf{73.56}\%  & 72.00\%  \\
\hline
\end{tabular}
\centering
 $^1$Results for three search terms,  $^2$Results for five search terms
\end{threeparttable}
}
\vspace{-.5cm}
\end{table*}

\subsection{Performance Metrics} \label{pmetrics}
\textbf{Mean Average Precision at K (MAPK)}: \emph{Precision at K} calculates \emph{precision} at the occurrence of every relevant result in the ranked list. \emph{Average Precision at K (APK)} averages the \emph{precision at K} for all relevant results in the list for a search query. Thus \emph{Mean Average Precision at K (MAPK)} is calculated from the mean of \emph{average precision at K} for all queries in the dataset as follows:
\begin{equation*}\label{eq:avep}
\setlength\abovedisplayskip{0pt}
\setlength\belowdisplayskip{0pt}
APK=\frac{\sum_{k=1}^{D}P_{k}\times rel_{k}}{\left |S \right |},~~ MAPK=\frac{\sum_{q\epsilon Q}APK(q)}{\left |Q\right |}
\end{equation*}
%
Here, $rel_{k}$ denotes the relevance function of $k^{th}$ result in the ranked list,  $P_{k}$ denotes the precision at $k^{th}$ result, and $D$ refers to number of total results. $S$ is the solution set for a query, and $Q$ is the set of all queries.

\textbf{Mean Recall (MR)}: \emph{Recall} denotes the fraction of the solution set that is retrieved for a search query. \emph{Mean Recall} averages such measures for all queries in the dataset.

\subsection{Experimental Results}\label{sec:result}
We conduct experiments using 349 change tasks from two subject systems, and apply four different metrics--\emph{no. of tasks solved}, \emph{\% of tasks solved}, \emph{mean average precision} and \emph{mean recall} for performance evaluation.
We consider different sizes for the search query, and collect the performance details of our suggested queries for both subject systems  which are reported in Table \ref{table:result}.
From the table, we note that our queries perform the best when five terms are used for search. For example, they return relevant results for 111 (48.26\%) out of 230 change tasks from \texttt{Log4j}, and for 60 (50.42\%) out of 119 change tasks from \texttt{eclipse.jdt.debug}, which is promising.
Thus, on average, our queries retrieve 63.33\% of the solutions from the dataset with a \emph{mean average precision} of 57.16\% for each of the subject systems. We also conduct search using six query terms; however, we note that our queries perform relatively poor in that regard for both subject systems. 

We investigate whether the \emph{preprocessing} (Section \ref{sec:sengine}) of corpus files significantly influences the performance of our queries. We re-ran the experiments on the corpus without preprocessing, and did not experience any major performance degradation. Thus, the findings actually demonstrate the robustness of our suggested search terms for a change task. We also investigate whether a list of randomly chosen five search terms from the \emph{Summary} of change task is comparable to our suggested search terms given that our algorithm also emphasizes on \emph{Summary} terms (Section \ref{sec:searchterm}). We conducted experiments with such queries from both subject systems where we noted that the queries performed quite poorly. For example, the search engine returns relevant results for at most 54 (compared to 111) tasks  from \texttt{Log4j} and 39 (compared to 60) tasks  from \texttt{eclipse.jdt.debug}. 
Thus, the findings demonstrate that our proposed technique for search term suggestion is relatively more effective and more reliable.

\subsection{Comparison with an Existing Approach}\label{sec:comparison}
We compare with one of the latest and closely related state-of-the-art approaches-- \citet{kevic} using our dataset (Section \ref{sec:dataset}). \citet{kevic} propose a heuristic model for search term suggestion for a change task where they consider frequency (\ie\ \emph{tf-idf}), location (\ie\ \emph{inSumAndBody, isInMiddle}) and notation (\ie\ \emph{isCamelCase}) of the terms from the task description. They suggest three search terms as a search query whereas we find our technique performing the best with five search terms in a query. We thus consider both sizes for the queries in our experiments.
We implement their \emph{relevance model} in our working environment, collect the search queries for the change tasks, and evaluate them using the same search engine (Section \ref{sec:sengine}). 
From Table \ref{table:comparison}, we note that our queries are more effective compared to theirs for both subject systems. For example, their queries can retrieve relevant results for at most 65 (28.26\%) out of 230 change tasks from \texttt{Log4j} and 39 (32.77\%) out of 119 tasks from \texttt{eclipse.jdt.debug}.
In terms of \emph{precision} and \emph{recall}, we note that our queries also perform relatively better than theirs.

\section{Related Work}\label{sec:related}
There exist a number of studies in the literature that attempt to support developers in \emph{feature location} tasks with search queries. They adopt different lightweight heuristics \cite{kevic} and query reformulation or expansion strategies \cite{gayg, refoqus, shepherd}, and perform different query quality analyses \cite{qquality,qeffect,specificity} and mining activities \cite{kevicdict,ccmapping}. However, most of these approaches expect a developer to provide the initial search query which they can improve upon, and their main focus is on improving a given query for a change task. On the other hand, in this study, we propose a novel technique that suggests a list of suitable terms as an initial search query for the given task. From technical perspective, we adapt an established algorithm-- \emph{TextRank} from information retrieval domain that analyzes the relative importance of the terms from the task description using a graph-based technique, and suggests the most important terms as search terms. Besides, we perform a case study using 349 change tasks that demonstrates the potential of the adapted technique.

\vspace{-.2cm}
\section{Conclusion \& Future Work\vspace{-.1cm}}
\label{sec:conclusion}
To summarize, in this paper, we propose a novel \emph{TextRank}-based technique that automatically identifies and suggests search terms for software change tasks.
Experiments with 349 change tasks from two  subject systems show that our approach, on average, can return relevant results (\ie\ Java classes) for 49.34\% of the change tasks with 57.16\% \emph{mean average precision} and 63.33\% \emph{recall}.
Comparison with one of the latest and closely related state-of-the-art approaches also shows that our approach performs comparatively better in different performance metrics. 
While these preliminary findings demonstrate the high potential of the proposed approach, further validation with more subject systems of diverse varieties and change tasks is warranted. 

\vspace{-.3cm}

\bibliographystyle{plainnat}
\setlength{\bibsep}{0pt plus 0.3ex}
\scriptsize
\bibliography{sigproc}  
%
%
\end{document}